\documentstyle{aipproc}

\begin{document}
\title{$\Lambda$-Inflation and CMBR Anisotropy}

\author{Vladimir N. Lukash}

\address{
Astro Space Center of P.N.Lebedev Physical Institute\\ 
Profsoyuznaya 84/32, Moscow, 117810, Russia}

\maketitle

\begin{abstract}
We argue that a $\Lambda$-inflation model can ensure large relative 
contribution of cosmic gravity waves into the $\Delta T/T$
at COBE scale preserving at the same time a near scale-invariant spectrum of 
cosmological density perturbations favored by observational data 
($n_S\simeq 1$). High efficiency of these models to meet observational tests 
is discussed.
\end{abstract}

\section*{Introduction}
Observational reconstruction of the power spectrum of cosmological density 
perturbations is a key problem of the modern cosmology. It provided a dramatic 
challenge after detecting the primordial CMB anisotropy by DMR COBE as the 
signal found at $10^0$ appeared to be few times higher than the expected 
value of $\Delta T/T$ in the most simple CDM model considered for
many years as a 'standard' cosmological model.  
During recent years there were many proposals to improve sCDM
by adding cosmological constant, hot dark matter, or considering 
non-scale-invariant primordial spectra of density perturbations. 

Presumably, a more natural way to solve the $\Delta T/T$ problem is an 
introduction of the cosmic gravitational waves contributing into large-scale 
CMB anisotropy. An inflationary model providing a sufficient abundance of
gravity waves should preserve the near-scale-invariant spectrum of density
perturbations as indicated by LSS observations.

A simple model of such kind is $\Lambda$-inflation, an inflationary model with
an effective metastable $\Lambda$-term\cite{lm}. We briefly
recall the physical properties of $\Lambda$-inflation and then consider 
cosmological implications of one of the model with self-interaction.

%%%%%%%%%%%%%%%%%%%%%%%%%%%%%%%%%%%%%%%
\section*{Physical properties of $\Lambda$-inflation}
%%%%%%%%%%%%%%%%%%%%%%%%%%%%%%%%%%%%%%%
\begin{itemize}
\item 
$\Lambda$-inflation is a class of single scalar field models with a positive
residual potential energy at the local minimum point, $V_0 > 0$. The hybrid
inflation model (at the intermediate evolutionary stage) is a particular case
of $\Lambda$-inflation: a relation of hybrid to $\Lambda$-inflation is similar
to the relation of massive field to chaotic inflation. The chaotic inflation
is a measure-zero model in the family of $\Lambda$-inflation models.
\item
The scalar perturbation spectrum generated in $\Lambda$-inflation has a
non-power-law wing-like shape with a broad minimum at $k=k_{cr}$ where the 
slope is locally scale invariant ($n_S=1$); it is blue, $n_S > 1$ 
(red, $n_S < 1$), on short, $k>k_{cr}$ (large, $k<k_{cr}$), scales. 
The tensor perturbation spectrum remains always red with a maximum deviation 
(from the Harrison-Zel'dovich one, $n_T=0$) at the scale near the S-spectrum 
minimum.
\item
The cosmic gravity waves generated in $\Lambda$-inflation contribute
maximally to the Sachs-Wolfe $\Delta T/T$ anisotropy, 
(T/S)${}_{max}{}_\sim^< 10$, at scale where the S-spectrum is slightly red or
nearly HZ ($k{}_\sim^< k_{cr}$). T/S remains small ($\ll 1$) in both blue
($k > k_{cr}$) and far red ($k \ll k_{cr}$) S-spectrum regions.
\item
Three independent arbitrary parameters determine $\Lambda$-inflation model;
they can be T/S, $k_{cr}$ (the scale where  $n_S = 1$), and $\sqrt{V_0}$
(the T-spectrum amplitude at $k_{cr}$ scale). This gives a high capability
for fitting various observational tests in the dark matter cosmologies
based on $\Lambda$-inflation.
\item
A simple estimate allows to relate straightforwardly the important parameters
of observational and physical cosmology:
\begin{equation}
\frac {\rm T}{\rm S}\simeq 3r \simeq -6 n_T \simeq 12 \gamma \simeq
\left(\frac{H}{6\times 10^{13}{\rm Gev}}\right)^2.
\end{equation} 
The ratio between tensor and scalar power spectra, $r$, and the inflationary 
$\gamma$ and Hubble, $H$, parameters are estimated in scale where the T/S 
is measured. A large T/S is expected if $(V_0)^{1/4}\sim 10^{16}$ Gev.
\end{itemize}

%%%%%%%%%%%%%%%%%%%%%%%%%%%%%%%%%%%%%%%%%%%%%%%%%%%
\section*{$\Lambda$-inflation with self-interaction}
%%%%%%%%%%%%%%%%%%%%%%%%%%%%%%%%%%%%%%%%%%%%%%%%%%%
A general potential of $\Lambda$-inflation can be presented as follows:
\begin{equation}
V(\varphi) = V_0 + \sum_{\kappa=2}^{\kappa_{max}} \frac{\lambda_{\kappa}}
{\kappa}\varphi^{\kappa},
\end{equation}
where $\varphi$ is the inflaton scalar field, $V_0 > 0$ and
$\lambda_{\kappa}$  are constants, and  $\kappa=2, 3, 4,..$.

In the case of a massive inflaton ($\kappa = \kappa_{max} = 2$, $\lambda_2 
\equiv m^2 > 0$) T/S can be larger than unity only when the S-spectrum 
slope in the `blue' asymptote is very steep, $n_S^{blue} > 1.8$. To avoid such
a strong spectral bend on short scales ($k > k_{cr}$), we choose another 
simple version of $\Lambda$-inflation -- the case with self-interaction 
(to be called $\Lambda\lambda$-inflation):  
$\kappa = \kappa_{max} = 4$, $\lambda_4 \equiv \lambda > 0$.

The scalar field $\varphi$ drives an inflationary evolution if $\gamma \equiv
- \dot H/H^2 < 1$ where $H \simeq \sqrt{V/3}$ (we assume Planckian units $8\pi 
G = c =\hbar = 1$). This condition holds true for all values of $\varphi$ if
\begin{equation} 
c \equiv \frac 14 \varphi_{cr}^2 =\frac 12
\sqrt{\frac{V_0}{\lambda}} > 1, 
\end{equation} 
The gravitational perturbation spectra $q_k$ and $h_k$ generated in S and T modes, 
respectively, are as follows \cite{luk}, \cite{lm}:
\begin{equation} 
q_k = \frac H{2\pi\sqrt{2\gamma}} = \frac{\sqrt{2\lambda/3}}{\pi}
\left(c^2+x^2\right)^{3/4},\;\;\;\;
h_k = \frac{H}{\pi\sqrt{2}} = \frac{2c\sqrt{\lambda/3}}{\pi} 
\left ( 1 + \frac x{\sqrt{c^2 + x^2}} \right)^{-1/2}, 
\end{equation}  
where 
\begin{equation}
x = \ln \left[ \frac{k}{k_{cr}} \left(1+\left(\frac xc\right)^2\right)^{1/4} \left(
1+\frac x{\sqrt{c^2+x^2}}\right)^{2/3}\right]\simeq\ln(k/k_{cr}).
\end{equation}

%%%%%%%%%%%%%%%%%%%%%%%%%%%%%%%%%%%%%%%%%%%%%%%%
\section*{CDM cosmology from $\Lambda\lambda$-inflation}
%%%%%%%%%%%%%%%%%%%%%%%%%%%%%%%%%%%%%%%%%%%%%%%%
Let us consider the S-spectrum with CDM transfer function, $T(k)$, normalized both 
at the large-scale $\Delta T/T\vert_{10^0}$ (including the contribution from 
T mode) and the galaxy cluster abundance at $z = 0$, and find a family of the 
most realistic S-spectra $q_k$ produced in $\Lambda\lambda$-inflation.

In total, we have three parameters entering the $q_k$ spectrum: $\lambda$, $c$
and $k_{cr}$. Constraining them by two observational tests, we are actually 
left with only one free parameter (say, $k_{cr}$) which may be restricted 
elsewhere by other observations.

To demonstrate explicitly how the three parameters are mutually related, we 
first employ a simple analytic estimates for the $\sigma_8$ and $\Delta T/T$ 
tests to derive the key equation relating $c$ and $k_{cr}$, and then solve 
this equation numerically to obtain the range of interesting physical 
parameters \cite{mnras}.

Instead of taking the $\sigma$-integral numerically we can estimate the spectrum 
amplitude on cluster scale as
\begin{equation}
q_{k_1} \simeq 4.5\times 10^{-7}\frac{h^2 \sigma_8}{k_1^2 T(k_1)}\;.
\end{equation}
On the other hand, the spectrum amplitude on large scale ($k_2=k_{COBE}\simeq
10^{-3}h/\rm{Mpc}$) can be derived from $\Delta T/T$ due to the Sachs-Wolfe
effect:
\begin{equation}
\big\langle\left(\frac{\Delta T}{T}\right)^2\big\rangle_{10^0} = {\rm S} + 
{\rm T} \simeq 1.1 \times 10^{-10},\;\;\;\; {\rm S} = 0.04 \; \langle q^2
\rangle_{\rm COBE}.
\end{equation}
The relationship between the power spectrum at COBE scale and the variance of 
the $q$ potential averaged in $10^0$-angular-scale, involves an effective 
interval of the spectral wavelengths contributing to the latter:
\begin{equation}
\langle q^2\rangle_{COBE}^{1/2} = fq_{k_2}, \;\;\;\;
f^2\sim \ln \left( \frac{k_2}{k_{hor}} \right).
\end{equation}
To estimate T/S we will use the approximation formula (1) for $x_2 = x_{COBE}$:
\begin{equation}
\frac{{T}}{S} \simeq -6n_T = \frac{6}{\sqrt{c^2 + x^2_2}}\left( 1-
\frac{x_2}{\sqrt{c^2 + x^2_2}}\right).
\end{equation}

Evidently, both normalizations determine essentially the corresponding $q_k$
amplitudes at the locations of cluster ($k_1$) and COBE ($k_2$) scales.
Accordingly, the parameters $k_1$ and $f$ can slightly vary while changing
the local spectrum slopes at the respective wavelengths. However, when the
deviation of $q_k$ slope from HZ is small we can just identify the 
parameters $k_1$ and $f$ for CDM models with their sCDM values:
\[
k_1 \simeq 0.3h/\rm{Mpc},\;\;\;\; f\simeq 1.26.
\]

Finally, we now write down the two normalization conditions as follows (for
$h=0.5$ and negligible baryon abundance):
\[
q_{k_1} \simeq 3\times 10^{-5}\sigma_8,\;\;\;\;
q_{k_2}\simeq 4\times 10^{-5}\left( 1 + \frac{\rm T}{\rm S}
\right)^{-1/2}.
\]
Comparing it with the theoretical spectrum $q_k$ we get the key equation for 
the relationship between $c$ and $k_{cr}$,
\begin{equation}
\left( \frac{q_{k_1}}{q_{k_2}} \right)^2 \simeq
D\left( 1 +\frac{\rm T}{\rm S}\right) ,
\end{equation}

Eq.(10) has a clear physical meaning: the ratio of the S-spectral power at
cluster and COBE scales is proportional to $\sigma_8^2$ and inversly
proportional to the fraction of the scalar mode contributing to the 
large-scale temperature anisotropy variance, S/(S+T).

Eq.(10) provides quite a general constraint for the fundamental 
inflation spectra in a wide set of dark matter models using only two basic
measurement (the cluster abundance and large scale $\Delta T/T$). Actually, 
all information on the dark matter is contained in the $D$-coefficient which 
can be calculated 
using the same equation (10) for a simple particular spectrum preserving the 
given model parameters:
\begin{equation}
D\simeq \frac{0.6\sigma_8^2}{1-3.1\Omega_b}\; ,\;\;\;\;\;
\Omega_b <0.2.
\end{equation}

The solution of eq.(10) is found in the plane $x - c$ for $D = 0.2, 0.3, 0.4$
numerically. For the entire range $0.1<D<0.5$ it can be analytically 
approximated with a precision less that 10\% as follows \cite{mnras}:
\begin{equation}
\ln^2 \left(\frac{k_0}{k_{cr}}\right)\simeq E\left(c_0-c\right)\left(c
+c_1\right)\;,\;\;\;\; 2<c\leq c_0\,
\end{equation}
\[
E\simeq 1,\;\;\;\ln k_0\simeq 49D^2+1.3,\;\;\;
c_0\simeq 61D^2+6.2,\;\;\; c_1\simeq 44D^2+4.0.
\]
The tensor mode contribution is approximated similarly ($k_0$ and $k_{cr}$ are
measured in units $h$/Mpc):
\begin{equation}
\frac{\rm T}{\rm S}\simeq \frac{2.53-4.3D}{(\ln k_{cr}+4.65)^{2/3}}
+\frac{1}{3}\;.
\end{equation}

%%%%%%%%%%%%%%%%%%%%
\section*{Discussion}
%%%%%%%%%%%%%%%%%%%%
We have presented a $\Lambda$-inflation model predicting a near scale-invariant
spectrum of density perturbations and large amount of cosmic gravitational
waves. This model is consistent with COBE $\Delta T/T$ and cluster abundance data. 
The perturbation spectra depend on one free scale-parameter, $k_{cr}$, which can 
be found in further analysis by fitting other observational data.  At the location of 
$k_{cr}$, the S-spectrum transients smoothly from the red ($k<k_{cr}$) to 
the blue ($k>k_{cr}$) parts.

Recent observational data favor a slightly red S-spectra at large scales
(cluster power spectrum, BOOMERanG and MAXIMA-1),
and require a blue increase of the spectrum at galaxy and subgalactic scales
($Ly_\alpha$ forest).   
This can be achieved by adjusting $k_{cr}$ to a cluster scale ($\sim 10\,h^{-1}$ 
Mpc).
 
%%%%%%%%%%%%%%%%%%%%%%%%%%
\section*{Acknowledgments}
%%%%%%%%%%%%%%%%%%%%%%%%%%
The work was partly supported by INTAS grant (97-1192). 
The author is grateful to the Organizing Committee for the hospitality and
financial support.

%%%%%%%%%%%%%%%%%

\end{document}